\documentstyle[12pt]{article}
\renewcommand{\theequation}{\arabic{equation}}

\newlength{\extraspace}
\newlength{\extraspaces}
\newcounter{dummy}

\newcommand{\be}{\begin{equation}
\addtolength{\abovedisplayskip}{\extraspaces}
\addtolength{\belowdisplayskip}{\extraspaces}
\addtolength{\abovedisplayshortskip}{\extraspace}
\addtolength{\belowdisplayshortskip}{\extraspace}}
\newcommand{\ee}{\end{equation}}

\newcommand{\ba}{\begin{eqnarray}
\addtolength{\abovedisplayskip}{\extraspaces}
\addtolength{\belowdisplayskip}{\extraspaces}
\addtolength{\abovedisplayshortskip}{\extraspace}
\addtolength{\belowdisplayshortskip}{\extraspace}}
\newcommand{\ea}{\end{eqnarray}}

\newcommand{\baa}{
\addtocounter{equation}{1}
\setcounter{dummy}{\value{equation}}
\setcounter{equation}{0}
\renewcommand{\theequation}{\arabic{dummy}\alph{equation}}
\begin{eqnarray}
\addtolength{\abovedisplayskip}{\extraspaces}
\addtolength{\belowdisplayskip}{\extraspaces}
\addtolength{\abovedisplayshortskip}{\extraspace}
\addtolength{\belowdisplayshortskip}{\extraspace}}
\newcommand{\eaa}{
\end{eqnarray}
\setcounter{equation}{\value{dummy}}
\renewcommand{\theequation}{\thesection.\arabic{equation}}}

\newcommand{\ban}{\begin{eqnarray*}
\addtolength{\abovedisplayskip}{\extraspaces}
\addtolength{\belowdisplayskip}{\extraspaces}
\addtolength{\abovedisplayshortskip}{\extraspace}
\addtolength{\belowdisplayshortskip}{\extraspace}}
\newcommand{\ean}{\end{eqnarray*}}

\newcommand{\ie}{{\it i.e.}}

\newcommand{\ra}{\rightarrow}

\newcommand{\Tau}{v}
\newcommand{\Time}{u}

\newcommand{\const}{c\,}
\newcommand{\UU}{\,\mbox{\large $\cal U$}}

\begin{document}
\addtolength{\baselineskip}{.3mm}
\input epsf

\vspace{.5cm}

\begin{center}
{\large\sc{Black Hole Evaporation 
and Complementarity\footnote{based on work with Y. Kiem, K. Schoutens
and H. Verlinde \cite{SVV,KVV}. This is a modified version of lecture
notes that will appear
in the proceedings of the Trieste Spring School of 1993.}}}\\[5mm]
{\it summary of lectures presented by Erik
Verlinde\footnote{TH-Division, CERN,
CH-1211 Geneva 23, Switzerland
and
Institute for Theoretical Physics,
University of Utrecht,
P.O. BOX 80.006, 3508 TA Utrecht, the Netherlands}
}\\[3mm]

\end{center}

\vspace{-1mm}
\noindent

About  twenty years ago  Hawking made the remarkable suggestion that
the black hole evaporation process will inevitably
lead to a fundamental loss of quantum coherence \cite{hawking}.
The mechanism by which  the quantum radiation is emitted appears
to be insensitive to the detailed history of the black hole, and thus
it seems
that most of the initial information is lost   for an outside observer.
 However, direct examination of Hawking's original derivation (or any
later one) of
the black hole emission spectrum shows that one inevitably needs
to make reference to particle waves that have arbitrarily high
frequency
near the horizon  as measured in the reference frame of the in-falling
matter.
This exponential red-shift effect associated with the black hole
horizon
leads to a breakdown of the usual separation of length scales  (see
e.g. \cite{thooft} and \cite{jacobson}), and effectively works as a
magnifying
glass that makes the consequences of the short distance, or rather,
high
energy physics near the horizon visible at larger scales to an
asymptotic observer.

Let us begin by reviewing the derivation of Hawking radiation in the
$s$-wave sector, while at first ignoring the effects due to
gravitational
back-reaction. We introduce two null-coordinates $u$ and
$v$  such that at a large distance $r\ra\infty$  we have $u\ra
t+r$ and $v\ra t-r$.  Following \cite{hawking}, we imagine sending a
small test-particle backwards in time from
future null infinity ${\cal I}^+$ and letting it propagate all the
way through to ${\cal I}^-$ (see figure 1.).
To relate the form of this signal in the two asymptotic regions, we use
the wave equation on the {\it fixed} background geometry
of the collapsing black hole.
\begin{center}
\leavevmode
\epsfysize=5.5cm
\epsfbox{fig1.eps}
\noindent
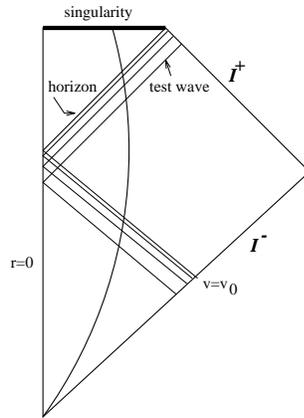
{\small Fig. 1}
\end{center}

{}From the condition that the field is regular at the origin $r=0$
one deduces that the outgoing $s$-wave $\phi_{out}$ of
the massless scalar field $\phi$ and the corresponding
in-coming wave $\phi_{in}$ are related by a reparametrization
\be\label{repar}
\phi_{in}(\Tau)=\phi_{out}(\Time(\Tau)),
\ee

\be
\label{ttau}
\Time(\Tau)=\Tau-4M\log[(\Tau_0-\Tau)/4M],
\ee
where $M$ denotes the black hole mass and $\Tau_0$ the critical
in-going
time, \ie\ the location of the in-going light-ray that later will
coincide
with the black hole horizon.
Thus an outgoing $s$-wave with a given frequency $\omega$ translates
to an in-signal
\be
\label{asym}
e^{i\omega \Time(\Tau)}=e^{i\omega\Tau} ({\Tau_0-\Tau\over
4M})^{-i4M\omega}
\theta(\Tau_0-\Tau),
\ee
that decomposes as a linear superposition of incoming waves
with very different frequencies.
 The out-going modes $b_\omega$ (\ie\ the Fourier coefficients of
$\phi_{out}$) are therefore related to the in-coming modes $a_\xi$
via a non-trivial Bogoljubov transformation of the form
\be
\label{Bogol}
b_\omega=\sum_{\xi} \alpha_{\omega\xi} a_\xi +
\sum_{\xi}\beta_{\omega\xi}
a^\dagger_\xi\quad;\qquad\quad
b^\dagger_\omega = \sum_{\xi} \beta^*_{\omega\xi} a_\xi
+ \sum_{\xi}\alpha^*_{\omega\xi}
a^\dagger_\xi.
\ee
where  $\alpha_{\omega\xi}$ and $\beta_{\omega\xi}$  are given by the
Fourier-coefficients of (\ref{asym}), for example
\be
\label{coeff}
\alpha_{\omega\xi} 
= \,e^{-i(\xi-\omega)v_0}  \,
{e^{2\pi M \omega}\Gamma(1-i4M\omega)\over2\pi\sqrt{\omega\xi}},
\ee

The transformation (\ref{Bogol}) is not invertible, and as a
consequence pure in-states
 $|in\rangle$ are mapped onto  {\it mixed} $out$-states. In particular,
the in-vacuum $|0\rangle$
evolves into
the famous Hawking state $\rho_H$, describing a constant flux of
thermal radiation
at the Hawking temperature $T_H= {1\over 8\pi M}$.

In the above description the incoming particles described by
$\phi_{in}(\Tau)$ with $\Tau >\Tau_0$ and the outgoing particles
represented by  $\phi_{out}(\Time)$ form independent
sectors of the Hilbert space,  because the corresponding field
operators commute with each
other.  The underlying classical intuition is that the fields
$\phi_{in}(\Tau)$
with $\Tau>\Tau_0$ will propagate into the region behind the black hole
horizon, and thus become unobservable from the out-side.
 However, here we have ignored the  fact that the in-falling
particles in fact {\it do} interact with the out-going radiation,
because they slightly change the
black hole geometry.   Furthermore, it can be seen from (\ref{asym})
that
 a generic out-going wave carries a diverging stress-energy near the
horizon.
 This indicates that gravitational interactions may indeed be
important.

Consider a spherical shell of  matter with energy $\delta M$ that
falls in to the black hole at some  late time $\Tau_1$.  At this time
the Schwarzschild radius  slightly  increases with an amount
$2\delta M$, and as a consequence the time $\Tau_0$ at which the
horizon
forms will also change very slightly (see figure 2). A simple
calculation shows that
\be
\label{deltatau}
\delta\Tau_0=-\const \delta M e^{-(\Tau_1-\Tau_0)/4M},
\ee
where, $c$ is a constant  of order one. At first it seems reasonable to
ignore this effect as long as the
change $\delta M$ is much smaller than $M$, but,   it turns out that,
due to the
diverging red-shift, the variation in $\Tau_0$,
although very small,  has an enormous effect on the wave-function
$\phi_{out}(\Time)$ of an out-going particle.
\begin{center}
\leavevmode
\epsfysize=8cm
\epsfbox{fig2.eps}
{\small Fig 2.}
\end{center}
\noindent
By combining (\ref{repar}), (\ref{ttau}) and (\ref{deltatau})
one easily verifies that as a result of the in-falling shell,
the outgoing particle-wave
is delayed by an amount that grows rapidly as a function of $\Time$
\be
\label{deltapsi}
\phi_{out}(\Time)\ra \phi_{out}(\Time-
4M\log(1-{\const} {\delta M \over 4M} e^{(\Time-\Tau_1)/4M})).
\ee
Notice that even for a very small perturbation $\delta M$  the argument
of
the field $\phi_{out}$  goes to infinity after a finite time
$\Time_{lim} -\Time_1 \sim -4M\log (\delta M/M)$.
The physical interpretation of this fact is
that a matter-particle that is on its way to reach the asymptotic
observer at some time $\Time>\Time_{lim}$ will,
as a result of the additional in-falling shell, cross the event-horizon
and
be trapped inside the
black-hole horizon.

To take the effect (\ref{deltapsi}) into account in the quantum theory,
 we divide up the  in-falling matter
in a classical piece plus a small quantum part that is described
in terms of a quantum field $\phi_{in}(\Tau)$.  Of course,
$\Tau_0$ is mainly determined by the classical in-falling matter, but
in
addition there is a small contribution that depends on the fields
$\phi_{in}(\Tau)$. For simplicity we assume here that the only
effect is that the mass and thus the Schwarschild radius changes as in
(\ref{deltatau}). In this way we find\footnote{For a
detailed discussion of the angular dependence we refer to \cite {KVV}.}
 \be
\label{newtau}
\Tau_0 =\Tau_0^{cl} - \const \int_{\Tau^{cl}_0}^\infty \!\! d\Tau \,
e^{(\Tau^{cl}_0-\Tau)/4M} T_{in}(\Tau)
\ee
where $T_{in}(\Tau)$ denotes the stress-energy tensor of the
$\phi_{in}(\Tau)$
with support $\Tau>\Tau^{cl}_0$.

With this correction we now re-calculate  the commutator of the
outgoing field
$\phi_{out}(\Time)$ for late times with the in-coming field
$\phi_{in}(\Tau)$ for $\Tau>\Tau^{cl}_0$.
{}For  this calculation we take  the same
relation (\ref{repar}), that formed the starting point of
Hawking's derivation, but we include in the reparametrization
(\ref{ttau}) the seemingly negligible quantum contribution in $\Tau_0$.
{}From the fact that the stress-tensor generates coordinate
transformations we deduce
\be
\lbrack \Tau_0,\phi_{in}(\Tau)\rbrack = -
i\const \exp((\Tau^{cl}_0-\Tau)/4M){\partial_\Tau}\phi_{in}(\Tau),
\ee
or equivalently
\be
\label{shift}
e^{i\xi\Tau_0}\phi_{in}(\Tau)e^{-i\xi\Tau_0}=\phi_{in}
(\Tau-4M\log(1-{\const\over 4M} \xi e^{(\Tau^{cl}_0-\Tau)/4M})),
\ee
Combining this result with  (\ref{repar}) and (\ref{ttau})
we obtain the following algebra for the $in-$ and $out$-fields
\be
\label{ex}
\phi_{out}(\Time)\phi_{in}(\Tau)=
\exp(i \const e^{(\Time-\Tau)/4M}\partial_\Time\partial_\Tau)
\phi_{in}(\Tau)\phi_{out}(\Time),
\ee
which is valid at for $\Tau>\Tau_0^{cl}$.
The `exchange algebra' (\ref{ex}) is the quantum
implementation of the gravitational back-reaction (\ref{deltapsi}) of
the
in-falling matter on the out-going radiation.   Notice that this
algebra is
non-local, and, furthermore, that the non-locality grows exponentially
with time.

These gravitational interactions lead to a modification of the quantum
state
of the out-going radiation. To describe this modification
we introduce addiditional modes $c_\omega$
describing the in-falling particles at the
horizon. In terms of these the original state computed by Hawking can
be written as:
\be
|\psi\rangle_{Hawking}
={1\over N} \exp\Bigl\lbrace
\int_0^\infty\!\!\!d\omega \,
e^{-4\pi M\omega} \, b^\dagger_{\omega} \, c^\dagger_{\omega }
\, \Bigr\rbrace \,
|0\rangle_b\otimes|0\rangle_c
\ee
But when one takes into account the correction (\ref{newtau})
due to the gravitational backreaction one obtains for final state
\be
\label{final}
|\psi\rangle_{final} =  \UU \, |\psi\rangle_{Hawking}
\ee
where
\be
\label{uudef}
\qquad \UU=
{\rm T} \exp \Bigl[\, i \! \int \!\!  \int_{v_0}\!\!\!
du dv \,e^{(u-v)/4M} T_{uu}(u) T_{vv}(v)\; \Bigr]
\ee
Notice again the interaction Hamiltonian wich appears in $\UU$ contains
a pre-factor that grows
exponentially with time. This makes clear that these interaction become
non-neglible at practically the same time when the black hole radiation
starts to appear.

We want to emphasize that in the above derivation we did not make any
assumptions other than those already made in the usual derivation of
Hawking evaporation.
The only extra ingredient that
we took into account is the small contribution to $\Tau_0$ due to the
field $\phi$. In this sense the relation (\ref{ex}) appears to be
unavoidable
and independent of which scenario one happens to believe in.
Therefore, it is clear that
the presence of these large commutators
implies that the standard semi-classical
picture of the black hole evaporation process needs to
be drastically revised. In particular, it tells us that, due
to the quantum uncertainty principle, we should be very careful in
making simultaneous statements about the in-falling and out-going
fields.

Our result supports the physical picture that there is
a certain {\it complementarity} between the physical realities
as seen by an asymptotic observer and by an in-falling
observer. Indeed, for the latter, the in-falling
matter will simply propagate freely without any perturbation, but he
(or she) will not see the out-going radiation. For the asymptotic
observer, on the other hand, the Hawking radiation is physically real,
while the in-falling matter will appear to evaporate completely before
it
falls into the hole.
Finally, we expect that our result may also be of importance
in relation with the entropy of black holes.
In \cite{thent} it was noted that a naive free field calculation of the
one-loop correction to the black hole entropy gives an infinite
answer. This infinity arises due to the diverging contribution
of states that are packed arbitrarily close to the horizon.
Our result suggests a possible remedy of this problem, because
it shows that in the coordinate system appropriate for this calculation
the in- and out-going fields no longer commute when they come
very close to the horizon. We believe that this
will effectively reduce the number of allowed states, and thereby
eliminate the diverging contribution in the entropy
calculation.

\renewcommand{\Large}{\large}

\medskip

\noindent
{\bf Acknowledgements.}

\noindent
The research of E.V. is partly supported by an Alfred P. Sloan
Fellowship,
and by a Fellowship of the Royal Dutch Academy of Sciences.

\end{document}